\documentclass[12pt]{article}

\usepackage{epsfig}

\newcommand{\re}{\mathrm{Re}\,}
\newcommand{\im}{\mathrm{Im}\,}

\newcommand{\gev}{\,{\rm GeV}}
\newcommand{\mev}{\,{\rm MeV}}

\newcommand{\half}{\textstyle \frac{1}{2}}

\newcommand{\mth}{m_\Theta}

\newcommand{\eth}{\eta_\Theta}

\begin{document}
\begin{flushright}
  CPHT-RR-112.1203 \\
  DESY-03-200 \\
  hep-ph/0312125
\end{flushright}
\vspace{\baselineskip}
\begin{center}
\textbf{\Large 
Probing the partonic structure of pentaquarks \\[0.3\baselineskip]
in hard electroproduction} \\
\vspace{2\baselineskip}
{\large M. Diehl$^{a}$,\, B.~Pire$^b$,\, L.~Szymanowski$^c$} \\
\vspace{3\baselineskip}
${}^a$\,Deutsches Elektronen-Synchroton DESY, 22603 Hamburg, Germany
\\[0.5\baselineskip]
${}^b$\,CPHT, {\'E}cole Polytechnique, 91128 Palaiseau, France
\\[0.5\baselineskip]
${}^c$\,Soltan Institute for Nuclear Studies, Ho\.{z}a 69, 00-681
Warsaw, Poland \\[0.5\baselineskip]
\vspace{4\baselineskip}
\textbf{Abstract}\\
\vspace{1\baselineskip}
\parbox{0.9\textwidth} {Exclusive electroproduction of a $K$ or $K^*$
meson on the nucleon can give a $\Theta^+$ pentaquark in the final
state.  This reaction offers an opportunity to investigate the
structure of pentaquark baryons at parton level.  We discuss the
generalized parton distributions for the $N \to \Theta^+$ transition
and give the leading order amplitude for electroproduction in the
Bjorken regime.  Different production channels contain complementary
information about the distribution of partons in a pentaquark compared
with their distribution in the nucleon.  Measurement of these
processes may thus provide deeper insight into the very nature of
pentaquarks.}
\end{center}
\vspace{1\baselineskip}
%


\section{Introduction}

There is increasing experimental evidence
\cite{Nakano:2003qx,Barth:2003es} for the existence of a narrow baryon
resonance $\Theta^+$ with strangeness $S=+1$, whose minimal quark
content is $uudd\bar{s}$.  Triggered by the prediction of its mass and
width in \cite{Diakonov:1997mm}, the observation of this hadron
promises to shed new light on our picture of baryons in QCD, with
theoretical approaches as different as the soliton picture
\cite{Diakonov:1997mm,Praszalowicz:2003ik}, quark models
\cite{Karliner:2003sy}, and lattice calculations \cite{Csikor:2003ng}, to
cite only a fraction of the literature.  A fundamental question is how
the structure of baryons manifests itself in terms of the basic
degrees of freedom in QCD, at the level of partons.  This structure at
short distances can be probed in hard exclusive scattering processes,
where it is encoded in generalized parton distributions
\cite{Muller:1994fv} (see \cite{Goeke:2001tz,Diehl:2003ny} for recent
reviews).  In this letter we introduce the transition GPDs from the
nucleon to the $\Theta^+$ and investigate electroproduction processes
where they could be measured, hopefully already in existing
experiments at DESY and Jefferson Lab.

In the next section we give some basics of the processes we propose to
study.  We then define the generalized parton distributions for the
$N\to \Theta$ transition and discuss their physics content (throughout
this paper we write $N$ for the nucleon and $\Theta$ for the
$\Theta^+$).  The scattering amplitudes and cross sections for
different production channels are given in Section~\ref{sec:scatter}.
In Section~\ref{sec:pole} we evaluate the contribution from kaon
exchange in the $t$-channel to the processes under study.  Concluding
remarks are given in Section~\ref{sec:concl}.


\section{Processes}
\label{sec:channels}

We consider the electroproduction processes
\begin{equation}
  \label{proc-p}
e p\to e \bar{K}^0 \, \Theta  , \qquad \qquad
e p\to e \bar{K}^{*0} \, \Theta ,
\end{equation}
where the $\Theta$ subsequently decays into $K^0 p$ or $K^+ n$.  Note
that the decay $\bar{K}^{*0} \to K^- \pi^+$ of the ${K}^{*}(892)$ tags
the strangeness of the produced baryon.  In contrast, the observation
of a $\bar{K}^0$ as $K_S$ or $K_L$ includes a background from final
states with a $K^0$ and an excited $\Sigma^+$ state in the mass region
of the $\Theta$, unless the strangeness of the baryon is tagged by the
kaon in the decay mode $\Theta\to K^+ n$.  Apart from their different
experimental aspects the channels with $\bar{K}$ or $\bar{K}^*$
production are quite distinct in their dynamics, as we will see in
Section~\ref{sec:scatter}.  We will also investigate the channels
\begin{equation}
  \label{proc-n}
e n\to e {K}^- \, \Theta  , \qquad \qquad
e n\to e {K}^{*-} \, \Theta
\end{equation}
accessible in scattering on nuclear targets.  The reconstruction of
the final state and of its kinematics is more involved in this case
because of the spectator nucleons in the target, but we will see in
Section~\ref{sec:scatter} that comparison of the processes
(\ref{proc-p}) and (\ref{proc-n}) may give valuable clues on the
dynamics.  We remark that the crossed process $K^+ n\to e^+e^-\,
\Theta$ could be analyzed along the lines of \cite{Berger:2001zn} at
an intense kaon beam facility.

\begin{figure}
\begin{center}
\leavevmode
\psfig{figure=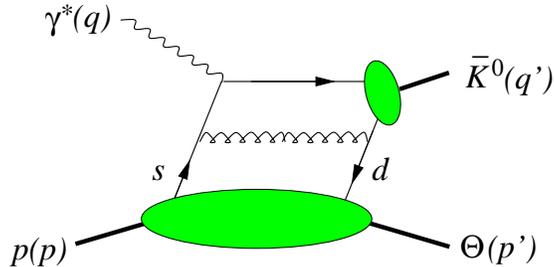,width=0.45\textwidth}
\end{center}
\caption{\label{fig:meson} One of the graphs for the $\gamma^* p \to
\bar{K}^0 \, \Theta$ amplitude in the Bjorken limit.  The large blob
denotes the GPD for the $p\to \Theta$ transition and the small one the
DA of the kaon.}
\end{figure}

The kinematics of the $\gamma^* p$ or $\gamma^* n$ subprocess is
specified by the invariants
\begin{equation}
  \label{invariants}
Q^2 = - q^2 , \qquad
W^2 = (p+q)^2 , \qquad
t = (p-p')^2 ,
\end{equation}
with four-momenta as given in Fig.~\ref{fig:meson}.  We are interested
in the Bjorken limit of large $Q^2$ at fixed $t$ and fixed scaling
variable $x_B = Q^2 /(2 pq)$.

According to the factorization theorem for meson production
\cite{Collins:1997fb}, the Bjorken limit implies factorization of the
$\gamma^* p$ amplitude into a perturbatively calculable subprocess at
quark level, the distribution amplitude (DA) of the produced meson,
and a generalized parton distribution (GPD) describing the transition
from $p$ to $\Theta$ (see Fig.~\ref{fig:meson}).  The dominant
polarization of the photon and (if applicable) the produced meson is
then longitudinal, and the corresponding $\gamma^* p$ cross section
scales like $d\sigma_L /(dt) \sim Q^{-6}$ at fixed $x_B$ and $t$, up
to logarithmic corrections in $Q^2$ due to perturbative evolution.  

We remark that pentaquarks with strangeness $S=-2$, like the
$\Xi^{--}$ recently reported in \cite{Alt:2003vb}, cannot be produced
from the nucleon by this leading-twist mechanism.  We also note that
if the $\Theta$ had isospin $I=2$ as proposed in
\cite{Capstick:2003iq} (but not favored by the experimental analyses
in \cite{Barth:2003es}), leading-twist electroproduction would be
isospin violating and hence tiny.

\begin{figure}
\begin{center}
\leavevmode
\psfig{figure=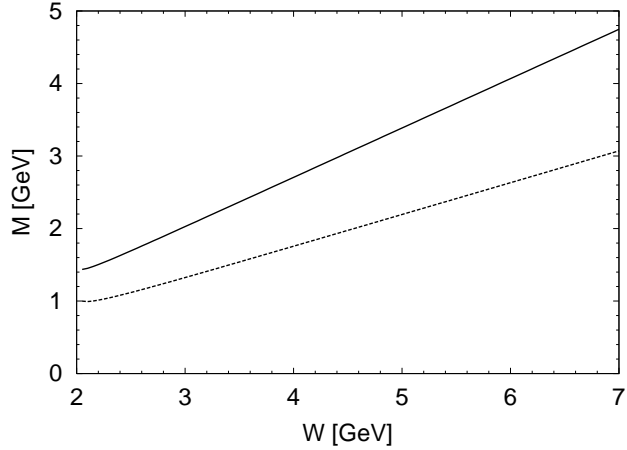,width=0.55\textwidth}
\end{center}
\caption{\label{fig:mass} Minimum invariant mass $M$ of $\bar{K}^0
K^0$ (lower curve) and of $\bar{K}^0 p$ (upper curve) in the process
$e p \to e \bar{K}^0 K^0 p$ at fixed invariant mass $1540 \mev$ of the
$K^0 p$ system.}
\end{figure}

The Bjorken limit implies a large invariant mass $W$ of the hadronic
final state, so that the produced baryon and meson are well separated
in phase space.  This provides a clean environment to study the
$\Theta$ resonance, with a low background obtained of course at the
price of a lower cross section than for inclusive production.  Large
enough $W$ in particular drives one away from kinematic reflections
which could fake a $\Theta$ resonance signal, discussed in
\cite{Dzierba:2003cm} for the process at hand.  To illustrate this we
show in Fig.~\ref{fig:mass} the smallest kinematically possible
invariant masses of the $\bar{K}^0 K^0$ and of the $\bar{K}^0 p$
system in $e p\to \bar{K}^0 K^0 p$ with the $K^0 p$ invariant mass
fixed at $\mth$.  Here and in the following we take $\mth = 1540 \mev$
in numerical evaluations (our results do not change significantly if
we take $\mth = 1525 \mev$ or $\mth = 1555 \mev$ instead).  We also
remark that strong interactions (in particular resonance) effects
between the $\bar{K}^0$ and the $K^0 p$ system will have a faster
power falloff than $Q^{-6}$ in the $\gamma^* p$ cross section at fixed
$x_B$, provided $Q^2$ is large enough for the analysis of the
factorization theorem to apply.

\section{The transition GPDs and their physics}
\label{sec:gpds}

Let us take a closer look at the transition GPDs that occur in the
processes we are interested in.  For their definition we introduce
light-cone coordinates $v^\pm = (v^0 \pm v^3) /\sqrt{2}$ and
transverse components $v_T = (v^1, v^2)$ for any four-vector $v$.  The
skewness variable $\xi = (p-p')^+ /(p+p')^+$ describes the loss of
plus-momentum of the incident nucleon and is connected with $x_B$ by
\begin{equation}
  \label{xi-vs-xB}
\xi \approx \frac{x_B}{2-x_B}
\end{equation}
in the Bjorken limit.

In the following we assume that the $\Theta$ has spin $J=\half$ and
isospin $I=0$.  Different theoretical approaches predict either $\eth
= 1$ or $\eth = -1$ for the intrinsic parity of the $\Theta$, and we
will give our discussion for the two cases in parallel.  The hadronic
matrix elements that occur in the electroproduction processes
(\ref{proc-p}) at leading-twist accuracy are
\begin{eqnarray}
  \label{matrix-elements}
F_V &=&
\frac{1}{2} \int \frac{d z^-}{2\pi}\, e^{ix P^+ z^-}
  \langle \Theta|\, \bar{d}(-\half z)\, \gamma^+ s(\half z) 
  \,|p \rangle \Big|_{z^+=0,\, {z}_T=0} \; ,
\nonumber \\
F_A &=&
\frac{1}{2} \int \frac{d z^-}{2\pi}\, e^{ix P^+ z^-}
  \langle \Theta|\, 
     \bar{d}(-\half z)\, \gamma^+ \gamma_5\, s(\half z)
  \,|p \rangle \Big|_{z^+=0,\, {z}_T=0}
\end{eqnarray}
with $P = \half (p+p')$, where here and in the following we do not
explicitly label the hadron spin degrees of freedom.  We define the
corresponding $p\to \Theta$ transition GPDs by
\begin{eqnarray}
  \label{gpd-pos}
F_V &=& \frac{1}{2P^+} \left[
  H(x,\xi,t)\, \bar{u}(p') \gamma^+ u(p) +
  E(x,\xi,t)\, \bar{u}(p') 
                 \frac{i \sigma^{+\alpha} (p'-p)_\alpha}{\mth+m_N} u(p)
  \, \right] ,
\nonumber \\
F_A &=& \frac{1}{2P^+} \left[
  \tilde{H}(x,\xi,t)\, \bar{u}(p') \gamma^+ \gamma_5 u(p) +
  \tilde{E}(x,\xi,t)\, \bar{u}(p') 
	\frac{\gamma_5\, (p'-p)^+}{\mth+m_N} u(p)
  \, \right]
\end{eqnarray}
for $\eth = 1$ and by
\begin{eqnarray}
  \label{gpd-neg}
F_V &=& \frac{1}{2P^+} \left[
  \tilde{H}(x,\xi,t)\, \bar{u}(p') \gamma^+ \gamma_5 u(p) +
  \tilde{E}(x,\xi,t)\, \bar{u}(p') 
	\frac{\gamma_5\, (p'-p)^+}{\mth+m_N} u(p)
  \, \right] ,
\nonumber \\
F_A &=& \frac{1}{2P^+} \left[
  H(x,\xi,t)\, \bar{u}(p') \gamma^+ u(p) +
  E(x,\xi,t)\, \bar{u}(p') 
        \frac{i \sigma^{+\alpha} (p'-p)_\alpha}{\mth+m_N} u(p)
  \, \right]
\end{eqnarray}
for $\eth = -1$.  Notice that the tilde in our notation indicates the
dependence on the spin of the hadrons, not on the spin of the quarks.
The scale dependence of the matrix elements is governed by the
nonsinglet evolution equations for GPDs
\cite{Muller:1994fv,Blumlein:1997pi}, with the unpolarized evolution
kernels for $F_V$ and the polarized ones for $F_A$.  Isospin
invariance gives $\langle \Theta | \bar{d}_\alpha s_\beta | p\rangle =
- \langle \Theta | \bar{u}_\alpha s_\beta | n\rangle$, so that the
transition GPDs for $n \to \Theta$ and those for $p \to \Theta$ are
equal up to a global sign.  For simplicity we write $F_V$, $F_A$ and
$H$, $E$, $\tilde{H}$, $\tilde{E}$ without labels for the transition
$p\to \Theta$.

\begin{figure}
\begin{center}
\leavevmode
\psfig{figure=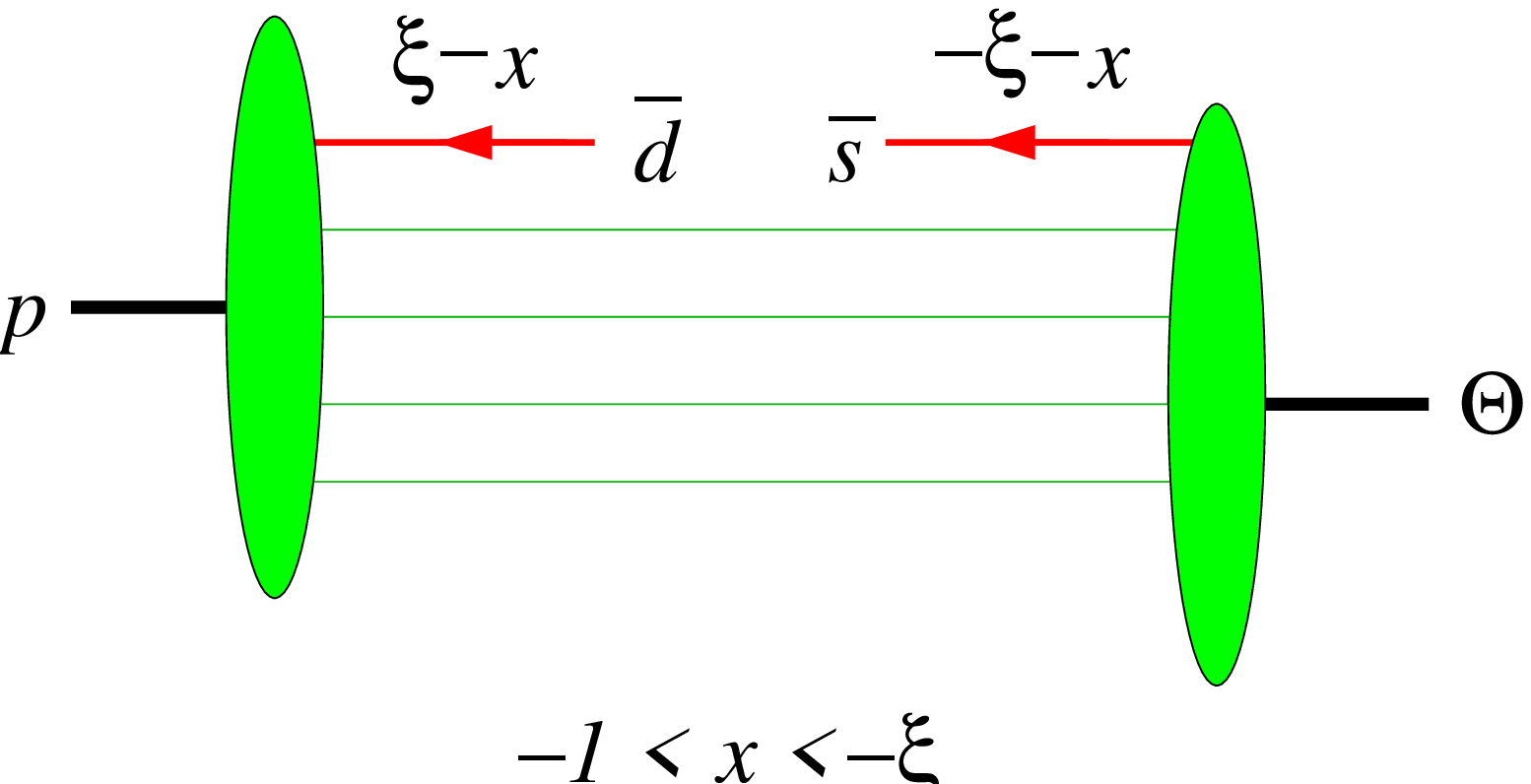,width=0.42\textwidth}
\hspace{2em}
\psfig{figure=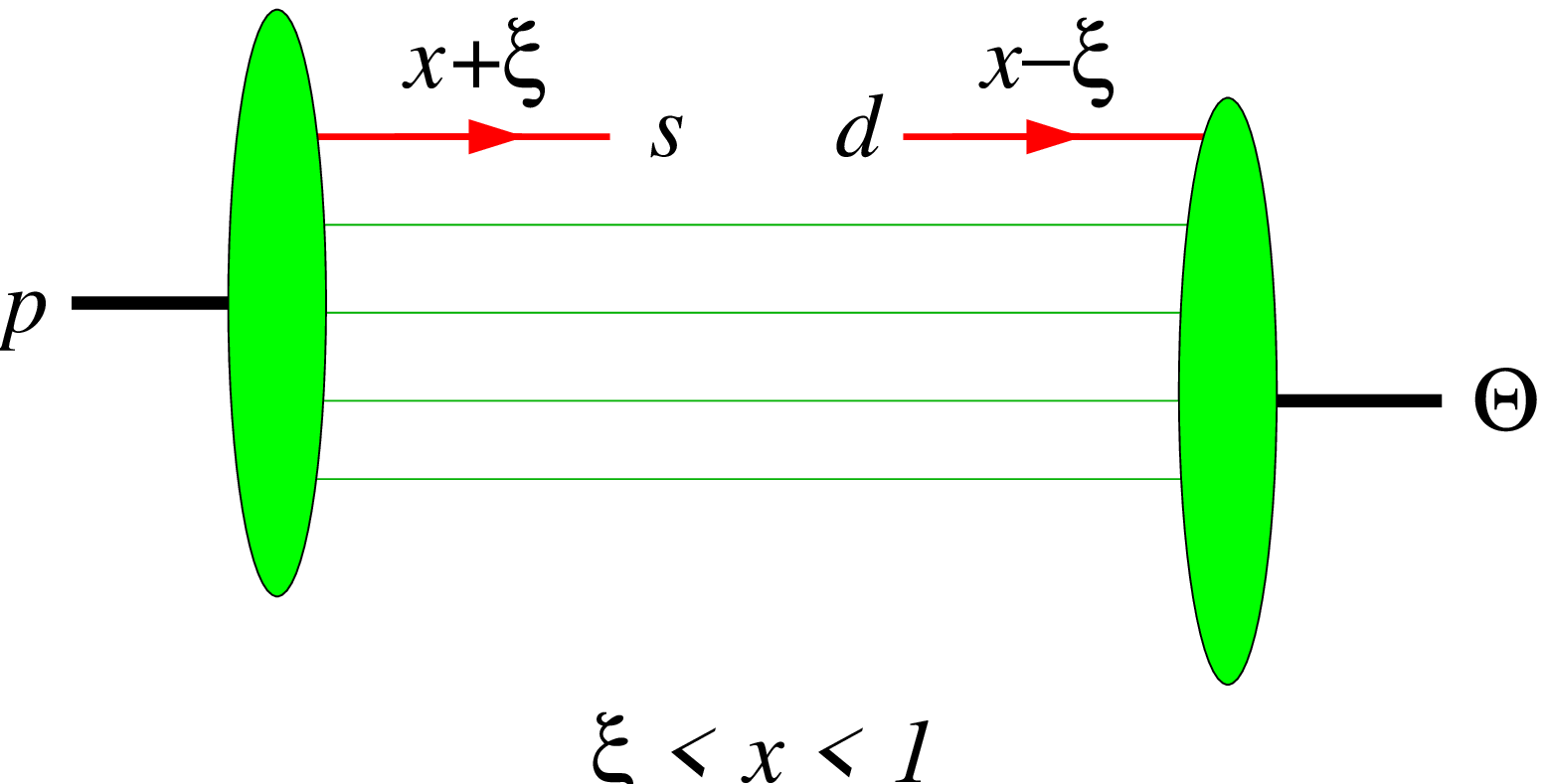,width=0.42\textwidth}

\vspace{2em}

\psfig{figure=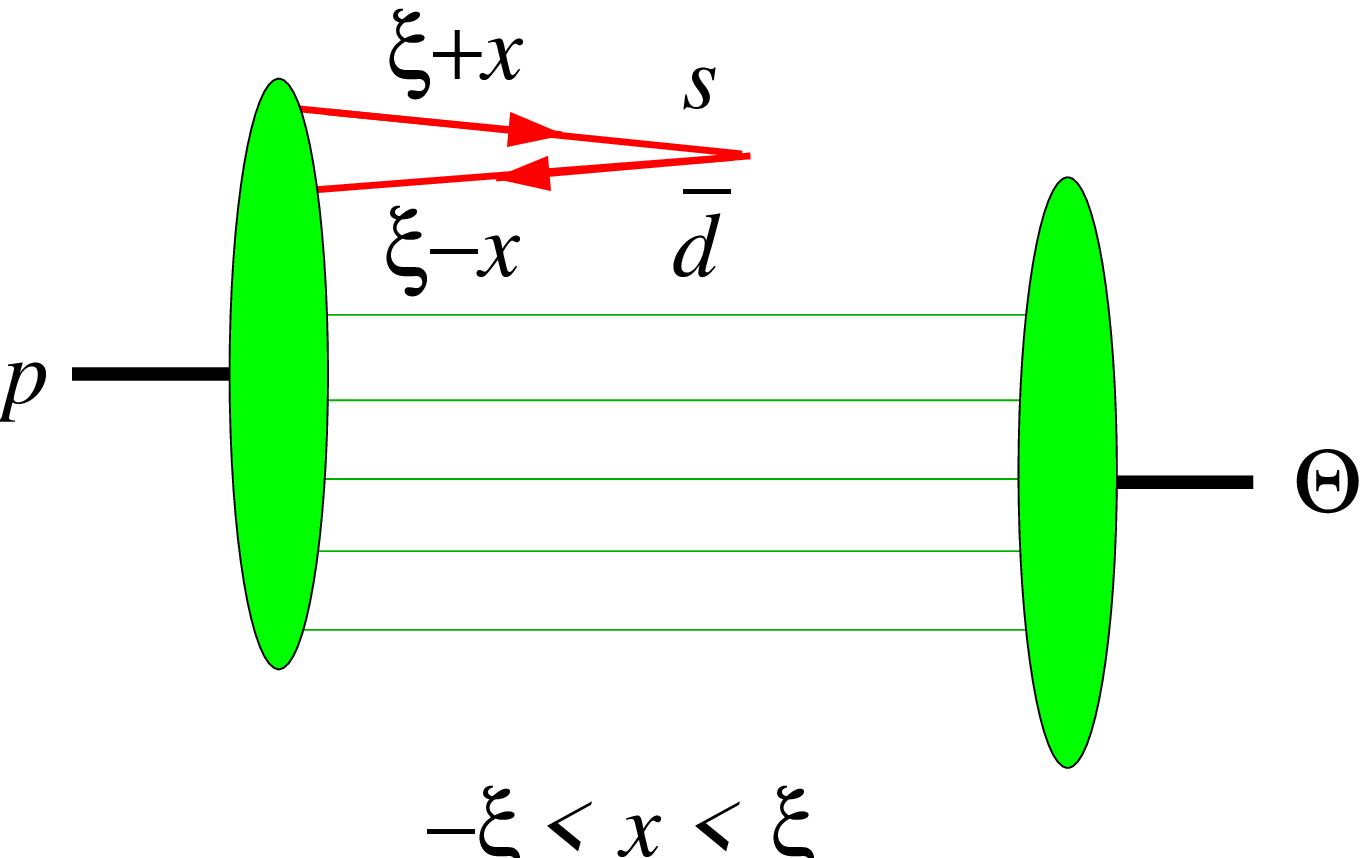,width=0.37\textwidth}
\end{center}
\caption{\label{fig:partons} Wave function representation of the
$p\to\Theta$ GPDs in the different regions of $x$.  The blobs denote
light-cone wave functions, and all possible configurations of
spectator partons have to be summed over.  The overall transverse
position of the $\Theta$ is shifted relative to the proton as
explained in \protect\cite{Diehl:2002he}.}
\end{figure}

The value of $x$ determines the partonic interpretation of the GPDs.
For $\xi<x<1$ the proton emits an $s$ quark and the $\Theta$ absorbs a
$d$ quark, whereas for $-1<x<-\xi$ the proton emits a $\bar{d}$ and
the $\Theta$ absorbs an $\bar{s}$.  The region $-\xi<x<\xi$ describes
emission of an $s\bar{d}$ pair by the proton.  In all three cases sea
quark degrees of freedom in the proton are involved.  The
interpretation of GPDs becomes yet more explicit when the GPDs are
expressed as the overlap of light-cone wave functions for the proton
and the $\Theta$.  As shown in Fig.~\ref{fig:partons}, the proton must
be in \emph{at least} a five-quark configuration for $\xi<|x|<1$ and
\emph{at least} a seven-quark configuration for $-\xi<x<\xi$.  We
emphasize however that all possible spectator configurations have to
be summed over in the wave function overlap, including Fock states
with additional partons in the nucleon and in the pentaquark.

As shown in \cite{Burkardt:2000za}, GPDs contain information about the
spatial structure of hadrons.  A Fourier transform converts their
dependence on $t$ into the distribution of quarks or antiquarks in the
plane transverse to their direction of motion in the infinite momentum
frame.  This tells us about the transverse size of the hadrons in
question.  The wave function overlap can also be formulated in this
impact parameter representation, with wave functions specifying
transverse position and plus-momentum fraction of each parton.  This
has in fact been done in Fig.~\ref{fig:partons}, and we refer to
\cite{Diehl:2002he} for a full discussion.  We see in particular that
for $\xi<|x|<1$ the transverse positions of all partons must match in
the proton and the $\Theta$, including the quark or antiquark taking
part in the hard scattering.  For $-\xi<x<\xi$ the transverse
positions of the spectator partons in the proton must match those in
the $\Theta$, whereas the $s$ and $\bar{d}$ are extracted from the
proton at the same transverse position (within an accuracy of order
$1/Q$ set by the factorization scale of the hard scattering process).
Note that small-size quark-antiquark pairs with net strangeness are
not necessarily rare in the proton, as is shown by the rather large
kaon pole contribution in the $p\to \Lambda$ transition (see the
discussion after (\ref{pole-factor}) below).  In summary, the $p\to
\Theta$ transition GPDs probe the partonic structure of the $\Theta$,
requiring the plus-momenta and transverse positions of its partons to
match with appropriate configurations in the nucleon.  The helicity
and color structure of the parton configurations must match as well.

We recall that for elastic transitions like $p\to p$ the analogs of
the matrix elements (\ref{matrix-elements}) reduce to the usual parton
densities in the forward limit of $\xi=0$ and $t=0$.  One then has
$H(x) = q(x)$, $H(-x) = -\bar{q}(x)$ and $\tilde{H}(x) = \Delta q(x)$,
$\tilde{H}(-x) = \Delta\bar{q}(x)$ for $x>0$, and the positivity of
parton densities results in inequalities like $| H(x) + H(-x) | \le
|H(x) - H(-x)|$ and $| \tilde{H}(x) | \le |H(x)|$.  One may expect
that this hierarchy persists at least in a limited region of nonzero
$\xi$ and $t$.  For the $p\to\Theta$ transition the situation is
different.  At given $\xi$ and $t$ the combinations $F_V(x) - F_V(-x)$
and $F_A(x) + F_A(-x)$ still give the sum of the configurations in
Fig.~\ref{fig:partons} with emission of a quark ($\xi<x<1$) and of an
antiquark ($\xi<-x<1$), whereas $F_V(x) + F_V(-x)$ and $F_A(x) -
F_A(-x)$ give their difference.  In the same $x$ regions $F_V$ still
gives the sum and $F_A$ the difference of configurations with positive
and negative helicity of the emitted and the absorbed parton.  There
are however no positivity constraints now, since the $p\to \Theta$
transition GPDs do not become densities in any limit.  They rather
describe the correlation between wave functions of $\Theta$ and
nucleon, which may be quite different.  Knowledge of the relative size
of the GPD combinations just discussed would in turn translate into
characteristic information about the wave functions of the $\Theta$
relative to those of the proton.

In the transition GPDs we have defined, the $\Theta$ is treated as a
stable hadron.  The amplitude of a full process, say $e p\to e
\bar{K}^0 K^0 p$ for definiteness, contains in addition a factor for
the decay $\Theta \to K^0 p$ and a term for the nonresonant $K^0 p$
continuum.  An alternative description is to use matrix elements
analogous to (\ref{matrix-elements}) directly for the hadronic state
$|K^0 p \rangle$ of given invariant mass, including both resonance and
continuum.  The leading-twist expression of the amplitude then
contains $p\to K^0 p$ transition GPDs, which have complex phases
describing the strong interactions in the $K^0 p$ system.  In the
partial wave relevant for the $\Theta$ resonance, these phases will
show a strong variation in the invariant $K^0 p$ mass around $\mth$.


\section{Scattering amplitude and cross section}
\label{sec:scatter}

The scattering amplitude for longitudinal polarization of photon and
meson at leading order in $1/Q$ and in $\alpha_s$ readily follows from
the general expressions for meson production given in
\cite{Diehl:2003ny}.  One has
\begin{eqnarray}
  \label{amp-p}
\mathcal{A}_{\gamma^* p\to \bar{K}^0\, \Theta} &=&
i e\, \frac{8\pi\alpha_s}{27}\, \frac{f_K}{Q}\, \Bigg[
I_K \int_{-1}^1 \frac{dx}{\xi-x-i\epsilon}\, 
	\Big( F_A(x,\xi,t) - F_A(-x,\xi,t) \Big)
\nonumber \\
 && \hspace{4.2em} {}+
J_K \int_{-1}^1 \frac{dx}{\xi-x-i\epsilon}\,  
	\Big( F_A(x,\xi,t) + F_A(-x,\xi,t) \Big)
\, \Bigg] ,
\nonumber \\
\mathcal{A}_{\gamma^* p\to \bar{K}^{*0}\, \Theta} &=&
i e\, \frac{8\pi\alpha_s}{27}\, \frac{f_{K^*}}{Q}\, \Bigg[
I_{K^*} \int_{-1}^1 \frac{dx}{\xi-x-i\epsilon}\,
	\Big( F_V(x,\xi,t) - F_V(-x,\xi,t) \Big)
\nonumber \\
 && \hspace{3.8em} {}+
J_{K^*} \int_{-1}^1 \frac{dx}{\xi-x-i\epsilon}\,  
	\Big( F_V(x,\xi,t) + F_V(-x,\xi,t) \Big)
\, \Bigg] ,
\end{eqnarray}
independently of the parity of the $\Theta$.  Our phase conventions
for meson states are fixed by
\begin{eqnarray}
  \label{kaon-decay}
\langle \bar{K}^0(q') 
	| \bar{s}(0) \gamma^\mu\gamma_5\, d(0) | 0\rangle
&=& \langle K^-(q') 
	| \bar{s}(0) \gamma^\mu\gamma_5\, u(0) | 0\rangle
\;=\; -i q'^\mu f_K ,
\nonumber \\
\langle \bar{K}^{*0}(q',\epsilon') 
	| \bar{s}(0) \gamma^\mu d(0) | 0\rangle
&=& \langle K^{*-}(q',\epsilon') 
	| \bar{s}(0) \gamma^\mu u(0) | 0\rangle
\;=\; -i \epsilon'^\mu m_{K^*} f_{K^*} ,
\end{eqnarray}
where $f_{K} = 160 \mev$, $f_{K^*} = (218 \pm 4) \mev$
\cite{Beneke:2003zv}, and $\epsilon'$ is the polarization vector of
the $K^*$.  This differs from the convention in \cite{Diehl:2003ny} by
the factors of $-i$ on the r.h.s.  In (\ref{amp-p}) we have integrals
\begin{eqnarray}
  \label{DA-integrals}
I &=& \int_0^1 dz\, \frac{1}{z(1-z)}\, \phi(z) 
	\;=\; 6 \sum_{n=0}^{\infty} a_{2n} ,
\nonumber \\
J &=& \int_0^1 dz\, \frac{2z-1}{z(1-z)}\, \phi(z)
	\;=\; 6 \sum_{n=0}^{\infty} a_{2n+1} ,
\end{eqnarray}
over the twist-two distribution amplitudes of either $\bar{K}^{0}$ or
$\bar{K}^{*0}$.  Our DAs are normalized to $\int_0^1 dz\, \phi(z) =
1$, and $z$ denotes the momentum fraction of the $s$-quark in the
kaon.  Because of isospin invariance $\bar{K}^0$ and $K^-$ have the
same DA, as have $\bar{K}^{*0}$ and $K^{*-}$.  In (\ref{DA-integrals})
we have used the expansion of DAs on Gegenbauer polynomials,
\begin{equation}
  \label{Gegenbauer}
\phi(z) = 6 z(1-z) \sum_{n=0}^\infty a_n \, C^{3/2}_n(2z-1)
\end{equation}
with $a_0=1$ due to our normalization condition.  Note that odd
Gegenbauer coefficients $a_{2n+1}$ are nonzero due to the breaking of
flavor SU(3) symmetry.  A recent estimate from QCD sum rules by Ball
and Boglione \cite{Ball:2003sc} obtained $a_1^{K^-} = -0.18 \pm 0.09$,
$a_2^{K^-} = 0.16 \pm 0.10$ and $a_1^{K^{*-}} = -0.4 \pm 0.2$,
$a_2^{K^{*-}} = 0.09 \pm 0.05$ at a factorization scale $\mu = 1
\gev$.  Note that the sign of $a_1$ in both cases is such that the $s$
quark tends to carry less momentum than the light antiquark, see the
discussion in \cite{Ball:2003sc}.  In contrast, Bolz et
al.~\cite{Bolz:1997ez} estimated $a_1^{K^-}$ to be of order $+0.1$ for
the kaon, using results of a calculation in the Nambu-Jona-Lasinio
model.

Note that the combination of GPDs going with $I_K$ corresponds to the
difference of quark and antiquark configurations in the sense of our
discussion at the end of Section~\ref{sec:gpds}.  In contrast, the
combination going with $I_{K^*}$ corresponds to the sum of quark and
antiquark contributions.  Given our ignorance about the relative sign
of the transition GPDs at $x$ and $-x$ we cannot readily say whether
the terms with $I$ or with $J$ tend to dominate in the amplitudes
(\ref{amp-p}).

For a neutron target the scattering amplitudes read
\begin{eqnarray}
  \label{amp-n}
\mathcal{A}_{\gamma^* n\to {K}^-\, \Theta} &=&
-i e\, \frac{8\pi\alpha_s}{27}\, \frac{f_K}{Q}\, \Bigg[
I_K \int_{-1}^1 \frac{dx}{\xi-x-i\epsilon}\, 
	\Big(F_A(x,\xi,t) + 2 F_A(-x,\xi,t) \Big)
\nonumber \\
 && \hspace{5.0em} {}+
J_K \int_{-1}^1 \frac{dx}{\xi-x-i\epsilon}\,  
	\Big(F_A(x,\xi,t) - 2 F_A(-x,\xi,t) \Big)
\, \Bigg],
\nonumber \\
\mathcal{A}_{\gamma^* n\to {K}^{*-}\, \Theta} &=&
-i e\, \frac{8\pi\alpha_s}{27}\, \frac{f_{K^*}}{Q}\, \Bigg[
I_{K^*} \int_{-1}^1 \frac{dx}{\xi-x-i\epsilon}\, 
	\Big(F_V(x,\xi,t) + 2 F_V(-x,\xi,t) \Big)
\nonumber \\
 && \hspace{4.6em} {}+
J_{K^*} \int_{-1}^1 \frac{dx}{\xi-x-i\epsilon}\,  
	\Big(F_V(x,\xi,t) - 2 F_V(-x,\xi,t) \Big)
\, \Bigg],
\end{eqnarray}
where we have used the isospin relations between the GPDs for $p\to
\Theta$ and $n\to \Theta$ and between the DAs for neutral and charged
kaons.  Due to the different factors for a photon coupling to $d$ and
$u$ quarks, the proton and neutron amplitudes involve different
combinations of GPDs at $x$ and $-x$.  Information on the relative
size of these combinations can thus be obtained by comparing data for
proton and neutron targets, given our at least qualitative knowledge
about the relative size of the integrals $I$ and $J$ over meson DAs.
If for example one had $F_A(x,\xi,t) \approx F_A(-x,\xi,t)$, the
amplitude for $\gamma^* p\to \bar{K}^0 \Theta$ would be dominated by
the SU(3) breaking integral $J_K$ and hence be suppressed, whereas no
such suppression would occur in the amplitude for $\gamma^* n \to K^-
\Theta$.  Comparison of $K$ and $K^*$ production on a given target can
in turn reveal the relative size between the matrix elements $F_A$ and
$F_V$.

\begin{figure}[b]
\begin{center}
\leavevmode
\psfig{figure=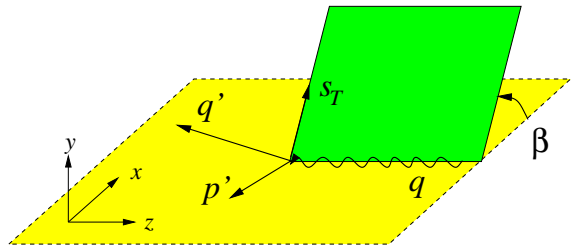,width=0.47\textwidth}
\end{center}
\caption{\label{fig:angle} Definition of the azimuthal angle $\beta$
between the hadronic plane and the transverse target spin $s_T$ in the
target rest frame.  $s_T$ is perpendicular to the $z$-axis, which
points in the direction opposite to the virtual photon momentum.}
\end{figure}

To leading accuracy in $1/Q^2$ and in $\alpha_s$ the cross section for
$\gamma^* p$ for a longitudinal photon on transversely polarized
target is
\begin{eqnarray}
  \label{X-section}
\frac{d\sigma_L}{dt} &=& 
\frac{64\pi^2 \alpha_{\mathit{em}}^{\phantom{2}} \alpha_s^2}{729}\, 
\frac{f_{K^{(*)}}^2}{Q^6}\, \frac{\xi^2}{1-\xi^2}\,
( S_U + S_T\, \sin\beta ) ,
\end{eqnarray}
where we use Hand's convention \cite{Hand:1963bb} for the virtual
photon flux.  $\beta$ is the azimuthal angle between the hadronic
plane and the transverse target spin as defined in
Fig.~\ref{fig:angle}.\footnote{Our convention for $\beta$ differs from
the one in \protect\cite{Goeke:2001tz,Frankfurt:1999fp}, with
$(\sin\beta)_{\mathrm{here}} = - (\sin\beta)_{[8],[22]}$.}
The cross section for an unpolarized target is simply obtained by
omitting the $\beta$-dependent term.  To have concise expressions for
$S_U$ and $S_T$ we define
\begin{eqnarray}
  \label{gpd-integrals}
\mathcal{H}(\xi,t) &=& 
I_{K^{(*)}} \int_{-1}^1 \frac{dx}{\xi-x-i\epsilon}\,
	\Big( H(x,\xi,t) - H(-x,\xi,t) \Big)
\nonumber \\
&+& J_{K^{(*)}} \int_{-1}^1 \frac{dx}{\xi-x-i\epsilon}\,
	\Big( H(x,\xi,t) + H(-x,\xi,t) \Big)
\end{eqnarray}
and analogous expressions $\mathcal{E}$, $\tilde\mathcal{H}$,
$\tilde\mathcal{E}$ for the other GPDs.  For $\eth = 1$ we have
\begin{eqnarray}
  \label{H-tilde-combinations}
S_U &=& (1-\xi^2) |\tilde\mathcal{H}|^2 
	+ \frac{(\mth-m_N)^2 - t}{(\mth+m_N)^2}\,
	  \xi^2 |\tilde\mathcal{E}|^2
	- \Bigg( \xi + \frac{\mth-m_N}{\mth+m_N} \Bigg)
	2\xi\, \re( \tilde\mathcal{E}^* \tilde\mathcal{H} ) \, ,
\nonumber \\
S_T &=& -\sqrt{1-\xi^2}\, \frac{\sqrt{t_0-t}}{\mth+m_N} \, 
	2\xi\, \im( \tilde\mathcal{E}^* \tilde\mathcal{H} )
\end{eqnarray}
for $K$ production and
\begin{eqnarray}
  \label{H-combinations}
S_U &=& (1-\xi^2) |\mathcal{H}|^2 
	- \Bigg( \frac{2\xi (\mth^2 - m_N^2) + t}{(\mth+m_N)^2} 
	+ \xi^2 \Bigg) |\mathcal{E}|^2
	- \Bigg( \xi + \frac{\mth-m_N}{\mth+m_N} \Bigg)
	2\xi\, \re( \mathcal{E}^* \mathcal{H} ) \, ,
\nonumber \\
S_T &=& \sqrt{1-\xi^2}\, \frac{\sqrt{t_0-t}}{\mth+m_N} \, 
	2\im( \mathcal{E}^* \mathcal{H} )
\end{eqnarray}
for $K^*$ production.  If $\eth = -1$ then
(\ref{H-tilde-combinations}) describes $K^*$ production and
(\ref{H-combinations}) describes $K$ production.  We see that one
cannot determine the parity of the $\Theta$ from the leading twist
cross section (\ref{X-section}) without knowledge about the dependence
of $\mathcal{H}$, $\mathcal{E}$, $\tilde\mathcal{H}$,
$\tilde\mathcal{E}$ on $t$ or $\xi$.  The same holds for scattering on
a neutron target, where one has to replace $H(-x,\xi,t)$ with $-2
H(-x,\xi,t)$ in (\ref{gpd-integrals}) and likewise change the
expressions for the other GPDs, as follows from (\ref{amp-p}) and
(\ref{amp-n}).

There is theoretical and phenomenological evidence that higher-order
corrections in $\alpha_s$ and in $1/Q$ can be substantial in meson
electroproduction at moderate values of $Q^2$, see
\cite{Goeke:2001tz,Diehl:2003ny} for a discussion and references.  For
$K^*$ production one can in particular expect an important
contribution from transverse polarization of the photon and the meson,
in analogy to what has been measured for exclusive electroproduction
of a $\rho^0$.  A minimum requirement for the applicability of a
leading-twist description is that $Q^2$ should be large compared to
$-t$ and $m^2_K$ or $m_{K^*}^2$, which directly enter in the
kinematics of the hard scattering process and should be negligible
there.  In kinematic relations, the squared baryon masses $m^2_N$ and
$\mth^2$ typically occur as corrections to terms of size $W^2$,
although a complete analysis of target mass corrections in exclusive
processes has not been performed yet.

There are arguments \cite{Frankfurt:1999fp,Goeke:2001tz,Diehl:2003ny}
that theoretical uncertainties from some of the corrections just
discussed cancel at least partially in suitable ratios of cross
sections.  At the level of the leading order formulae (\ref{amp-p})
and (\ref{amp-n}) we see for instance that the scale uncertainty in
$\alpha_s$ cancels in the ratio of cross sections on a proton and a
neutron target, and that the dependence on the meson structure comes
only via the ratio $J/I$.  Other processes to compare with are given
by $ep \to e K^0 \Sigma^+$, $ep \to e K^+ \Sigma^0$, $ep \to e K^+
\Lambda$ or their analogs for vector kaons or a neutron target, with
the production of either ground state or excited hyperons.  Such
channels may also be useful for cross checks of experimental
resolution and energy calibration.  Their amplitudes are given as in
(\ref{amp-p}) with an appropriate replacement of matrix elements $F_V$
or $F_A$ listed in Table~\ref{tab:channels}.  We have used isospin
invariance to replace the transition GPDs from the neutron with those
from the proton.  Isospin invariance further gives $F_{p\to \Sigma^+}
= \sqrt{2}\, F_{p\to \Sigma^0}$.

\begin{table}
\caption{\label{tab:channels} Combinations of transition GPDs
multiplying $I$ and $J$ in the hard scattering formula
(\protect\ref{amp-p}) and its analogs for the listed channels.}
$$
\renewcommand{\arraystretch}{1.2}
\begin{array}{lll} \hline\hline
  & ~~~~~~~~~~~~~~~I  & ~~~~~~~~~~~~~~~J  \\ \hline
\gamma^* p \to \bar{K}^0 \Theta &  
\phantom{-} F_{p\to \Theta}(x) - F_{p\to \Theta}(-x) &
\phantom{-} F_{p\to \Theta}(x) + F_{p\to \Theta}(-x) \\
\gamma^* p \to K^0 \Sigma^+ &
\phantom{-} F_{p\to \Sigma^+}(x) - F_{p\to \Sigma^+}(-x) &
- [ F_{p\to \Sigma^+}(x) + F_{p\to \Sigma^+}(-x) ] \\
\gamma^* p \to K^+ \Sigma^0 &
- [ 2F_{p\to \Sigma^0}(x) + F_{p\to \Sigma^0}(-x) ] &
\phantom{-} 2F_{p\to \Sigma^0}(x) - F_{p\to \Sigma^0}(-x) \\
\gamma^* p \to K^+ \Lambda &
- [ 2F_{p\to \Lambda}(x) + F_{p\to \Lambda}(-x) ] &
\phantom{-} 2F_{p\to \Lambda}(x) - F_{p\to \Lambda}(-x) \\ \hline
\gamma^* n \to {K}^- \Theta &  
- [ F_{p\to \Theta}(x) + 2F_{p\to \Theta}(-x) ] &
- [ F_{p\to \Theta}(x) - 2F_{p\to \Theta}(-x) ] \\
\gamma^* n \to K^+ \Sigma^- &
\phantom{-} 2F_{p\to \Sigma^+}(x) + F_{p\to \Sigma^+}(-x) &
- [ 2F_{p\to \Sigma^+}(x) - F_{p\to \Sigma^+}(-x) ] \\
\gamma^* n \to K^0 \Sigma^0 &
- [ F_{p\to \Sigma^0}(x) - F_{p\to \Sigma^0}(-x) ] &
\phantom{-} F_{p\to \Sigma^0}(x) + F_{p\to \Sigma^0}(-x) \\
\gamma^* n \to K^0 \Lambda &
\phantom{-} F_{p\to \Lambda}(x) - F_{p\to \Lambda}(-x) &
- [ F_{p\to \Lambda}(x) + F_{p\to \Lambda}(-x) ] \\ \hline\hline
\end{array}
\renewcommand{\arraystretch}{1}
$$
\end{table}

For transitions within the ground state baryon octet, SU(3) flavor
symmetry relates the transition GPDs to the flavor diagonal ones for
$u$, $d$ and $s$ quarks in the proton \cite{Frankfurt:1999fp},
\begin{eqnarray}
F_{p\to \Lambda} &=& \frac{1}{\sqrt{6}}\,
	\Big( F^s_{p\to p} + F^d_{p\to p} - 2F^u_{p\to p} \Big) ,
\nonumber \\
F_{p\to \Sigma^0} &=&  \frac{1}{\sqrt{2}}\,
	\Big( F^s_{p\to p} - F^d_{p\to p} \Big) .
\end{eqnarray}
One may expect these relations to hold reasonably well, except for the
distributions $\tilde{E}$, where SU(3) symmetry is strongly broken by
the difference between pion and kaon mass in the respective pole
contributions (see the following section).  In the approximation of
SU(3) symmetry, comparison of $\Theta$ production with the
corresponding hyperon channels would thus compare the $N\to \Theta$
transition GPDs with the GPDs of the nucleon itself.


\section{Kaon pole contributions}
\label{sec:pole}

\begin{figure}
\begin{center}
\leavevmode
\psfig{figure=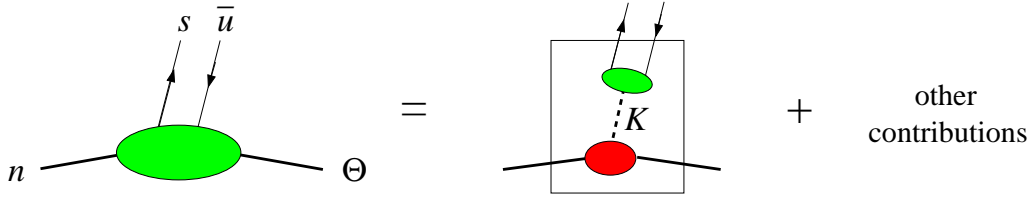,width=0.85\textwidth}
\end{center}
\caption{\label{fig:kaon-pole} Kaon pole contribution to the $n\to
\Theta$ transition GPDs in the region $-\xi<x<\xi$.}
\end{figure}

In analogy to the well-known pion exchange contribution to the elastic
nucleon GPDs, the axial vector matrix elements $F_A$ for the
transition between nonstrange and strange baryons receive a
contribution from kaon exchange in the $t$-channel, as shown in
Fig.~\ref{fig:kaon-pole}.  It can be expressed in terms of the kaon
distribution amplitude and the appropriate baryon-kaon coupling if
$t=m_K^2$.  This is of course outside the physical region for our
electroproduction processes, where the contribution from the kaon pole
is expected to be less and less dominant for increasing $-t$.  With
this caveat in mind we will now discuss the kaon pole contribution to
the $N\to \Theta$ GPDs, as this can be done without a particular
dynamical model for the~$\Theta$.

\begin{figure}
\begin{center}
\leavevmode
\psfig{figure=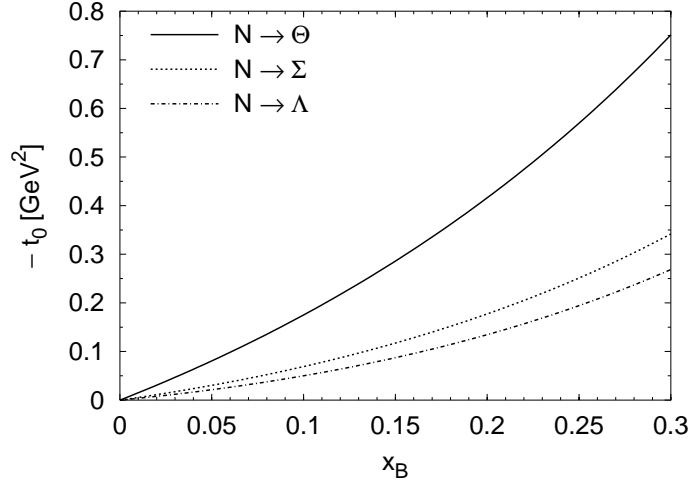,width=0.59\textwidth}
\end{center}
\caption{\label{fig:tmin} Minimum kinematically allowed value of $-t$
at given $x_B$ for the transitions $N\to \Theta$, $N\to \Sigma$, $N\to
\Lambda$, see (\protect\ref{xi-vs-xB}) and (\protect\ref{tmin}).}
\end{figure}

We recall at this point that the minimal kinematically allowed value
of $-t$ at given $\xi$,
\begin{equation}
  \label{tmin}
-t_0 = 
  \frac{2 \xi^2 (\mth^2 + m_N^2) + 2 \xi (\mth^2 - m_N^2)}{1-\xi^2} ,
\end{equation}
is not so small in typical kinematics of fixed target experiments.
This is shown in Fig.~\ref{fig:tmin}, where we have replaced $\xi$
with $x_B$ using the relation (\ref{xi-vs-xB}) valid in Bjorken
kinematics.  We also show the corresponding values of $-t_0$ for the
transition from the nucleon to a ground state $\Sigma$ or $\Lambda$.

We define the $\Theta NK$ coupling through
\begin{equation}
\mathcal{L} = ig_{\Theta NK} K_d (\bar{\Theta} \gamma_5 p)
- ig_{\Theta NK} K_u (\bar{\Theta} \gamma_5 n) + \textrm{c.c.}
\end{equation}
if $\eth = 1$, and through
\begin{equation}
  \label{g-def-negative}
\mathcal{L} = ig_{\Theta NK} K_d (\bar{\Theta} p)
- ig_{\Theta NK} K_u (\bar{\Theta} n) + \textrm{c.c.}
\end{equation}
if $\eth = -1$.  Here $K_d$ denotes the field that creates a
$\bar{K}^0$ and $K_u$ the one creating a $K^-$.  The factor of $i$ in
(\ref{g-def-negative}) is dictated by time reversal invariance, since
we choose the phase of the $\Theta$ field such that it has the same
transformation under time reversal as the nucleon field.  Then the
GPDs defined in (\ref{gpd-neg}) are real valued.  The above
definitions can be rewritten in terms of the vector or axial vector
current using the free Dirac equation for the $\Theta$ and the nucleon
fields.  Using the method of \cite{Mankiewicz:1998kg} we obtain kaon
pole contributions
\begin{eqnarray}
  \label{gpd-pole-pos}
\xi \tilde{E}_{\mathrm{pole}} &=&
	\frac{g_{\Theta NK} f_K (\mth+m_N)}{m_K^2 - t}\, 
	\frac{1}{2} \phi\Big(\frac{x+\xi}{2\xi}\Big)
\nonumber \\
\tilde{H}_{\mathrm{pole}} &=& H_{\mathrm{pole}} 
	\;=\; E_{\mathrm{pole}} \;=\; 0
\end{eqnarray}
for $\eth = 1$ and
\begin{eqnarray}
  \label{gpd-pole-neg}
E_{\mathrm{pole}} &=& -H_{\mathrm{pole}} =
	\frac{g_{\Theta NK} f_K (\mth+m_N)}{m_K^2 - t}\, 
	\frac{1}{2} \phi\Big(\frac{x+\xi}{2\xi}\Big)
\nonumber \\
\tilde{H}_{\mathrm{pole}} &=& \tilde{E}_{\mathrm{pole}} \;=\; 0
\end{eqnarray}
for $\eth = -1$, where it is understood that $x$ is limited to the
region between $-\xi$ and $\xi$, and where $\phi$ is the same kaon
distribution amplitude we have encountered earlier.  At the level of
the amplitudes (\ref{amp-p}) and (\ref{amp-n}) for $K$ production one
finds
\begin{eqnarray}
  \label{amplitude-pole}
\mathcal{A}_{\gamma^* p\to \bar{K}^0\, \Theta}^{\mathrm{pole}} &=& 
	ie\, \bar{u}(p') \gamma_5\, {u}(p)\, 
	\frac{g_{\Theta NK}}{m_K^2 - t}\, Q F_{\bar{K}^0}(Q^2) ,
\nonumber \\
\mathcal{A}_{\gamma^* n\to {K}^-\, \Theta}^{\mathrm{pole}} &=& 
	-ie\, \bar{u}(p') \gamma_5\, {u}(p)\,
	\frac{g_{\Theta NK}}{m_K^2 - t}\, Q F_{K^-}(Q^2)
\end{eqnarray}
for $\eth = 1$, whereas for $\eth = -1$ one simply has to replace
$\bar{u}(p') \gamma_5\, {u}(p)$ with $\bar{u}(p') {u}(p)$ in both
relations.  Here
\begin{eqnarray}
  \label{kaon-ff}
F_{\bar{K}^0}(Q^2) &=& - \frac{2\pi \alpha_s}{9}\, \frac{f_K^2}{Q^2}\,
  \frac{4}{3} I_K J_K
\nonumber \\
F_{K^-}(Q^2) &=& - \frac{2\pi \alpha_s}{9}\, \frac{f_K^2}{Q^2}\,
  \Big( I_K^2 - \frac{2}{3} I_K J_K + J_K^2 \Big)
\end{eqnarray}
are the elastic kaon form factors at leading accuracy in $1/Q^2$ and
$\alpha_s$.  We note that the relations (\ref{amplitude-pole}) remain
valid beyond this approximation, which in analogy to the pion form
factor we expect to receive important corrections at moderate $Q^2$,
see \cite{Diehl:2003ny} for references.  The form factors are
normalized as $F_{K^-}(0) = -1$ and $F_{\bar{K}^0}(0) = 0$, and at
nonzero $t$ the neutral kaon form factor is only nonzero thanks to
flavor SU(3) breaking.  The contribution of the squared kaon pole
amplitude to the $\gamma^* p \to \bar{K}^0 \Theta$ or $\gamma^* n \to
K^- \Theta$ cross section finally reads
\begin{equation}
  \label{X-section-pole}
\frac{d\sigma_L}{dt} \Bigg|_{\mathrm{pole}} =
\alpha_{\mathit{em}}^{\phantom{2}}
\frac{F_K^2(Q^2)}{Q^2}\, \frac{x_B^2}{4 (1-x_B)}\;
   g_{\Theta NK}^2 \, \frac{(\mth- \eth m_N)^2 - t}{(m_K^2-t)^2} ,
\end{equation}
where $F_K$ is the appropriate form factor for the $\bar{K}^0$ or the
$K^-$.  Of course, the pole contribution (\ref{amplitude-pole}) also
appears in the cross section via its interference with the non-pole
parts of the amplitude, which we cannot estimate at this point.

For kinematical reasons $\Theta\to K^0 p$ and $\Theta \to K^+ n$ are
the only strong decays of the $\Theta$, so that its total width
$\Gamma_\Theta$ translates to a good accuracy into a value of
$g^2_{\Theta NK}$,
\begin{eqnarray}
  \label{theta-decay}
\Gamma_\Theta = \frac{g^2_{\Theta NK}}{4\pi}\, 
  k\; \frac{(\mth- \eth m_N)^2 - m_K^2}{\mth^2} ,
\end{eqnarray}
where $k \approx 268 \mev$ is the momentum of the decay nucleon in the
$\Theta$ rest frame.  Taking an indicative value of $\Gamma_\Theta =
10 \mev$ we obtain $g_{\Theta NK}^2 /(4\pi) = 0.77$ for $\eth = 1$ and
$g_{\Theta NK}^2 /(4\pi) = 0.015$ for $\eth = -1$.  The squared
couplings corresponding to different values of $\Gamma_\Theta$ are
readily obtained by simple rescaling.

\begin{figure}
\begin{center}
\leavevmode 
\psfig{figure=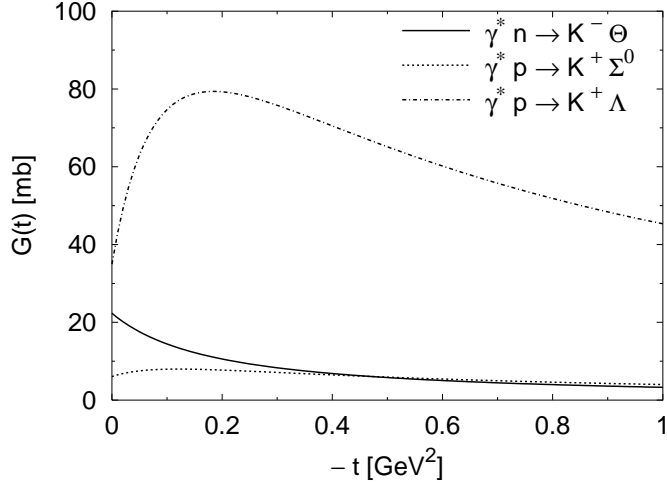,width=0.59\textwidth}
\end{center}
\caption{\label{fig:pole} The kaon pole factor $G(t)$ defined in
(\protect\ref{pole-factor}) and evaluated for $\Gamma_\Theta = 10
\mev$ and $\eth = 1$, and the analogous factors for the transitions $p
\to \Sigma^0$ and $p \to \Lambda$.  The same factors appear in the
kaon pole contributions to $p \to \Theta$, $n \to \Sigma^0$ and $n \to
\Lambda$.}
\end{figure}

To be insensitive to the theoretical uncertainties in evaluating the
kaon form factors, we compare in the following the kaon pole
contributions to different baryon transitions.  In Fig.~\ref{fig:pole}
we show the factor
\begin{equation}
  \label{pole-factor}
G(t) = g_{\Theta NK}^2 \, 
	\frac{(\mth- \eth m_N)^2 - t}{(m_K^2-t)^2}
\end{equation}
appearing in the pole contribution (\ref{X-section-pole}) to the
$\gamma^* n\to K^- \Theta$ cross section, as well as
its analogs for the pole contributions to $\gamma^* p \to K^+
\Sigma^0$ and to $\gamma^* p \to K^+ \Lambda$.  Due to isospin
invariance the corresponding factor for $\gamma^* n \to K^+ \Sigma^-$
is twice as large as for $\gamma^* p \to K^+ \Sigma^0$.  Following
\cite{Frankfurt:1999xe} we take $g_{\Sigma NK}^2 /(4\pi) = 1.2$ and
$g_{\Lambda NK}^2 /(4\pi) = 14$ for the couplings between the proton
and the neutral hyperons.  As an indication of their uncertainties one
may compare these values with those given in~\cite{Guidal:2003qs},
namely $g_{\Sigma NK}^2 /(4\pi) = 1.6$ and $g_{\Lambda NK}^2 /(4\pi) =
10.6$.  We remark that according to the estimates of
\cite{Frankfurt:1999xe}, the overall cross section for $\gamma^* p\to
K^+ \Lambda$ is comparable in size to the one for $\gamma^* p \to
\pi^+ n$ in kinematics where both processes receive substantial
contributions from the kaon or pion pole.

\begin{figure}
\begin{center}
\leavevmode
\psfig{figure=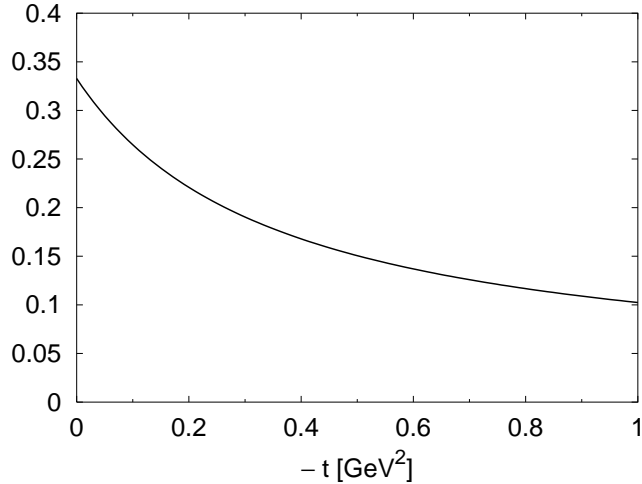,width=0.59\textwidth}
\end{center}
\caption{\label{fig:pole-ratio} Ratio of the kaon pole cross sections
(\protect\ref{X-section-pole}) for the cases $\eth = -1$ and $\eth =
1$ at given $\Gamma_\Theta$.}
\end{figure}

Note that the much smaller coupling for a negative-parity $\Theta$ is
partially compensated in the kaon pole contribution to the cross
section by a larger kinematic factor in the numerator of
(\ref{X-section-pole}).  The ratio of the factors $G(t)$ for $\eth=-1$
and for $\eth=1$ is shown in Fig.~\ref{fig:pole-ratio}.  Given the
presence of contributions not due to the kaon pole it is not clear
whether one could use the measured size and $t$ dependence of the
cross section to infer on the parity of the $\Theta$.

The factors $G(t)$ shown in Fig.~\ref{fig:pole} also describe the
neutral kaon pole contributions in $\gamma^* p \to \bar{K}^0 \Theta$,
$\gamma^* n \to K^0 \Sigma^0$ and $\gamma^* n \to K^0 \Lambda$.
Compared with the respective charged kaon pole contributions in
$\gamma^* n \to {K}^- \Theta$, $\gamma^* p \to K^+ \Sigma^0$ and
$\gamma^* p \to K^+ \Lambda$, they are significantly suppressed by a
factor $(F_{\bar{K}^0} /F_{K^-})^2$ at cross section level.  This
factor is about 0.03 if we take the leading-order expressions
(\ref{kaon-ff}) of the form factors together with the estimates of
\cite{Ball:2003sc} for the Gegenbauer coefficients $a_1$ and $a_2$ in
the kaon DA, given below (\ref{Gegenbauer}).


\section{Conclusions}
\label{sec:concl}

We have investigated exclusive electroproduction of a $\Theta^+$
pentaquark on the nucleon at large $Q^2$, large $W^2$ and small $t$.
Such a process provides a rather clean environment to study the
structure of pentaquark at parton level, in the form of well defined
hadronic matrix elements of quark vector or axial vector currents.  In
parton language, these matrix elements describe how well parton
configurations in the $\Theta$ match with appropriate configurations
in the nucleon (see Fig.~\ref{fig:partons}).  Their dependence on $t$
gives information about the size of the pentaquark.  Channels with
production of pseudoscalar or vector kaons and with a proton or
neutron target carry complementary information.  The transition to the
$\Theta$ requires sea quark degrees of freedom in the nucleon, and we
hope that theoretical approaches including such degrees of freedom
will be able to evaluate the matrix elements given in
(\ref{matrix-elements}).  Candidates for this may for instance be the
chiral quark-soliton model or lattice QCD, both of which have been
used to calculate the corresponding matrix elements for elastic
nucleon transitions, see \cite{Petrov:1998kf} and
\cite{Gockeler:2003jf}.

In order to obtain observably large cross sections one may be required
to go to rather modest values of $Q^2$, where the leading
approximation in powers of $1/Q^2$ and of $\alpha_s$ on which we based
our analysis receives considerable corrections.  The associated
theoretical uncertainties should be alleviated by comparing $\Theta$
production to the production of $\Sigma$ or $\Lambda$ hyperons as
reference channels.  In any case, even a qualitative picture of the
overall magnitude and relative size of the different hadronic matrix
elements accessible in the processes we propose would give information
about the structure of pentaquarks well beyond the little we presently
know about these intriguing members of the QCD spectrum.


\section*{Acknowledgments}  

We thank E. C. Aschenauer, M. Gar\c{c}on, D. Hasch, and G. van der
Steenhoven for helpful discussions.  The work of B. P. and L. Sz.\ is
partially supported by the French-Polish scientific agreement
Polonium.  CPHT is Unit{\'e} mixte C7644 du CNRS.


\end{document}